\documentclass[hyper]{JHEP3}
\usepackage{graphicx}
\usepackage{float}
\usepackage{amsmath}
\usepackage{amssymb}
\usepackage{subfig}
\usepackage[square, comma, sort&compress]{natbib}

\newcommand{\beq}{\begin{eqnarray}}
\newcommand{\eeq}{\end{eqnarray}}
\newcommand{\eq}{eqnarray}

\newcommand{\la}{{\lambda}}
\newcommand{\La}{{\Lambda}}

\newcommand{\om}{{\omega}}

\newcommand{\f}{\frac}
\newcommand{\de}{{\scriptsize \textsc{de}}}

\newcommand{\currentDE}{\Omega_{\rm DE}}

\title{Constraining the Detailed Balance Condition in Ho\v{r}ava Gravity
with Cosmic Accelerating Expansion}
\author{Chien-I Chiang$^{1, 2 \sharp}$, Je-An Gu$^{2 \dag}$, Pisin Chen$^{1, 2, 3, 4 \ddag}$
\\$^1$Department of Physics, National Taiwan University, Taipei 10617, Taiwan, R.O.C.,
\\$^2$Leung Center for Cosmology and Particle Astrophysics,
National Taiwan University, Taipei 10617, Taiwan, R.O.C.
\\$^3$Graduate Institute of Astrophysics, National Taiwan University, Taipei 10617, Taiwan, R.O.C.
\\$^4$Kavli Institute for Particle Astrophysics and Cosmology,
SLAC National Accelerator Laboratory, Menlo Park, CA 94025, U.S.A.
\\
\\
$^\sharp$\email{b95203016@ntu.edu.tw}\\
$^\dag$\email{jagu@ntu.edu.tw}\\
$^ \ddag$\email{pisinchen@phys.ntu.edu.tw}\\
}

\abstract{In 2009 Ho\v{r}ava proposed a power-counting renormalizable quantum gravity theory. Afterwards a term in the action that softly violates the detailed balance condition has been considered with the attempt of obtaining a more realistic theory in its IR-limit. This term is proportional to $\omega R^{(3)}$, where $\omega$ is a constant parameter and $R^{(3)}$ is the spatial Ricci scalar. In this paper we derive constraints on this IR-modified Ho\v{r}ava theory using the late-time cosmic accelerating expansion observations. We obtain a lower bound of $|\omega|$ that is nontrivial and depends on $\Lambda_W$, the cosmological constant of the three dimensional spatial action in the Ho\v{r}ava gravity. We find that to preserve the detailed balance condition, one needs to fine-tune $\Lambda_W$ such that $- 2.29\times 10^{-4}< (c^2 \Lambda_W)/(H^2_0 \currentDE) - 2 < 0 $, where $H_0$ and $\currentDE$ are the Hubble parameter and dark energy density fraction in the present epoch, respectively. On the other hand, if we do not insist on the detailed balance condition, then the valid region for $\Lambda_W$ is much relaxed to $-0.39< (c^2 \Lambda_W)/(H^2_0 \currentDE) - 2 < 0.12$. We find that although the detailed balance condition cannot be ruled out, it is strongly disfavored.}

\begin{document}
\section{Introduction}

In 2009 Ho\v{r}ava proposed a power-counting renormalizable quantum gravity theory, which is non-relativistic in the high-energy, or UV, limit and recovers the Lorentz symmetry in the low-energy, or IR, limit \cite{HoravaPRD09,HoravaJHEP09}. Various aspects of this
theory have been widely pursued
\cite{Cai:2009JHEP,Cai:2009PRD,Nishioka:2009CQG,Charmousis:2009JHEP,Sotiriou:2009JHEP,
Germani:2009JHEP,Kobakhidze:2009zr,Bogdanos:2009CQG,Kluson:2009JHEP,Afshordi:2009PRD,Myung:2009PLB,
Alexandre:2009PRD,Blas:2009,Capasso:2009JHEP,ChenBin:2009,Kluson:2009PRD,Kiritsis:2009PRD,
Garattini:2009,Kluson:2010,Son:2010,Carloni:2010,Eune:2010,Li:2009JHEP,Visser:2009PRD,
Blas:2009JHEP,Henneaux:2009PRD,Pons:2010ke,Bellorin:2010}.
In addition to the theory itself, its various implications in cosmology have also drawn
wide attention \cite{CalcagniJHEP09, KiritsisNPB09,Chaichian:2010yi,Elizalde:2010ep}. Specifically, there has been a large amount of effort in studying
the cosmological perturbations \cite{Gao:2009PRD,
Mukohyama:2009JCAP:Perturbation, Piao:2009PRB, Chen:2009JCAP,
Cai:2009PRD:Perturbation, Wang:2009PRD, Kobayashi:2009JCAP,
GaoXian:2009, Kobayashi:2010JCAP}, black hole properties
\cite{Kiritsis:2009JHEP, ParkJHEP09, Danielsson:2009JHEP,
Cai:2009PRD:BlackHole, Kehagias:2009PLB,
Myung:2009PLB:Thermo:BlackHole, BottaCantcheff:2009, Lee:2009rm,
Greenwald:2009PRD,Majhi:2009xh,Koutsoumbas:2010pt}, gravitational waves \cite{Takahashi:2009wc,Mukohyama:2009PLB,
Myung:2009PLB:GravitationalWave, Koh:2009, Park:2009}, dark energy
phenomenology \cite{ParkJCAP10, Saridakis:2009EPJ, Jamil:2010,
Ali:2010sv}, observational constraints on the parameters of the
theory \cite{Dutta:2009JCAP:Observation, Dutta:2010}, and many
others.

With regard to the background space-time expansion, Ho\v{r}ava gravity in principle modifies
the GR Friedmann equation with additional terms stemming from
its non-conventional gravity, thereby contributing to the dark sector.
Accordingly, the current and future cosmic observations may provide significant constraints on
Ho\v{r}ava gravity, especially when connecting it
with the cosmic accelerating expansion.

In this context it was found that the effective Friedmann equation derived from
Ho\v{r}ava gravity with the detailed balance condition would include a cosmological
constant (CC) term and a radiation-like $a^{-4}$ term
\cite{CalcagniJHEP09, KiritsisNPB09},
where $a$ is the scale factor in the Friedmann-Robertson-Walker
(FRW) metric. This radiation-like term originates from the
fourth-order spatial derivative terms in the Ho\v{r}ava action. On the other hand,
if the detailed balance condition is violated, then there would be other dark terms induced.

Although the gravity action of the Ho\v{r}ava theory with the detailed
balance condition recovers that of GR in the IR limit, its
solution may not be so. For example, in \cite{LuPRL09} it was
shown that the black hole solution does not recover the usual
AdS-Schwarzschild black hole solution in GR. In order to
attain a more desirable IR behavior without abandoning the
simplicity provided by the detailed balance condition in the UV
limit, several authors introduced a soft violation of the detailed
balance condition \cite{HoravaPRD09, Nastase09, KehagiasPLB09},
namely, a term proportional to the spatial three-curvature
$R^{(3)}$. In particular, an IR-modified Ho\v{r}ava theory that
accommodates flat Minkowski vacuum was studied in
\cite{KehagiasPLB09}. The exact solutions of spherical symmetry
with and without matter were obtained in \cite{Kim:2010}. \\

A question then naturally arises: to what extent can the Ho\v{r}ava gravity violate the detailed balance condition? One possible means to address this question would be to derive constraints on the IR-modification terms and those obeying the detailed balance
condition from cosmological observations. For this purpose we consider a cosmological model
studied in \cite{ParkJCAP10} based on the IR-modified Ho\v{r}ava gravity,
with the FRW metric describing the background space-time and with the
energy content that includes radiation and dust matter.

In \cite{ParkJCAP10} Park showed that the Friedmann equation of
this cosmological model contains additional $a^{-4}$, $a^{-2}$, and CC
terms beyond that in GR. Park identified these terms as the effective dark
energy (DE) that is responsible for the cosmic acceleration. The observations
about the expansion history can in principle constrain the behavior of the
(effective) dark energy and thereby constrain the Ho\v{r}ava gravity.
%
To constrain this Ho\v{r}ava Effective Dark Energy (\emph{HEDE})
by observations, an efficient approach is via a phenomenological
parametrization of the relevant physical quantities that have been well
studied. Once the relation between the model parameters
and the phenomenological parameters is established, the constraints on the model
can be obtained from those on the phenomenological parameters that have been derived from observations.
Park considered the widely used Chevallier-Polarski-Linder (CPL)
parameterization of the equation of state of dark energy
\cite{Chevallier00IJMP,LinderPRL03},
\begin{equation} \label{CPL}
w_\de (a) \equiv p_\de / \rho_\de = w_0 + w_a(1-a) ,
\end{equation}
where the constraints on the phenomenological parameters $w_0$
and $w_a$ from the updated observations have been well studied
(see, for example, \cite{IchikawaJCAP08,XiaPRD08,ChenPLB09,7yearWMAP1,7yearWMAP2}).

Park \cite{ParkJCAP10} explored the feasibility of \emph{HEDE} by
considering three best-fit values of $(w_0, w_a)$ obtained in
\cite{IchikawaJCAP08, XiaPRD08} where a non-flat universe was
considered. It was suggested by Park that the existence of some \emph{HEDE} models
that satisfy these three best-fit values indicates the validity of
\emph{HEDE}. We note, however, that in principle the dark
energy density of \emph{HEDE} is determined once the values of $(w_0, w_a)$ are
given. As will be shown in the present paper, these three best-fit
models predict the dark energy densities that are much smaller than that
required by observations, and have thus already been ruled out.

In the present paper we pursue a more comprehensive test of the
IR-modified Ho\v{r}ava theory and its resultant \emph{HEDE} based on
the current observations. We particularly emphasize that a
complete test of \emph{HEDE} based on the
cosmic expansion must take into consideration not only the
evolution of the dark energy density $\rho_\de(a)$, which involves
both the present value of $\currentDE$ and the equation of state
$w_\de(a)$, but also the present value of the
fractional density $\Omega_k$ of the spatial curvature.

Specifically, the present density fraction of dark energy
$\currentDE$ should be around $0.74$ and that of the spatial
curvature $\Omega_k<0.01$ \cite{7yearWMAP1}. In addition, since in
the dark sector the radiation-like term would be dominant in the
early universe, its energy density must be smaller than the true
radiation energy density $\Omega_r$, otherwise \emph{HEDE} would
predict a later epoch of the matter-radiation equality and that in turn would
be in conflict with the cosmic
microwave background (CMB) and the big-bang nucleosynthesis (BBN) observational results. As
will be shown in the present paper, the observational constraint on
the dark energy equation of state $w_\de$, together with the above
three observational requirements, very tightly constrains
\emph{HEDE}. These observational constraints on \emph{HEDE} favor
the violation of detailed balance of Ho\v{r}ava gravity. We will
present a lower bound on the extent of the violation.

This paper is organized as follows. In Sec.\ 2 we give a brief
review of \emph{HEDE}, a cosmological model based on the
IR-modified Ho\v{r}ava theory. In Sec.\ 3 we discuss the general
strategy that we utilize for the model test, which involves an approximate
relation between the model parameters and the phenomenological
parameters. In Sec.\ 4 we investigate the observational
constraints on \emph{HEDE} that are presented in the
phenomenological parameter space. In Sec.\ 5 we show how the
constraints on the phenomenological parameters are transcribed into
that on the model parameters, and investigate its impact on the
IR-modified Ho\v{r}ava gravity. In Sec.\ 6 we show the evolution
patterns of the effective dark energy in the \emph{HEDE} models
that are consistent with observations. We conclude in Sec.\ 7.

\section{Setup of the Model}

To be self-contained, in this section we give a brief review of
the IR-modified Ho\v{r}ava-gravity cosmological model that was
investigated by Park \cite{ParkJCAP10}. The action of the IR-modified
Ho\v{r}ava gravity reads
\begin{eqnarray}
    S_g & = & \displaystyle \int dt d^3x \sqrt{g} N
              \left[
              \frac{2}{\kappa^2} \left( K_{ij}K^{ij}-\lambda K^2\right)
              - \frac{\kappa^2}{2 \nu^4}C_{ij}C^{ij}
              + \frac{\kappa^2 \mu}{2\nu^2}\epsilon^{ijk}R^{(3)}_{il}\nabla_j {R^{(3)l}}_k
              \right.  \\
        &  & \left.
              - \frac{\kappa^2 \mu^2}{8}R^{(3)}_{ij}R^{(3)ij}
              + \frac{\kappa^2 \mu^2}{8(3 \lambda -1)}
              \left( \frac{4\lambda -1}{4} (R^{(3)})^2 - \Lambda_W R^{(3)} + 3 \Lambda^2_W\right)
              + \frac{\kappa^2 \mu^2 \omega}{8 (3\lambda -1)}R^{(3)}
              \right] \nonumber ,
\end{eqnarray}
where the extrinsic curvature
\begin{equation}
K_{ij}=\frac{1}{2N} \left(\dot{g}_{ij}-\nabla_i N_j-\nabla_jN_i\right)
\end{equation}
(the dot denotes the time derivative), the Cotton tensor
\begin{equation}
C^{ij}=\epsilon^{ik\ell}\nabla_k
\left(R^{(3)j}{}_\ell-\frac{1}{4}R^{(3)} \delta^j_\ell\right) ,
\end{equation}
and $\kappa,\lambda,\nu,\mu, \Lambda_W,\omega$ are constant
parameters. Note that on the right-hand side of Eq.(2.1) the last term proportional to $\omega R^{(3)}$ induces the soft violation of the detailed balance condition.
For a homogeneous and isotropic universe we consider a
FRW metric of the form
\begin{equation}
ds^2=-c^2dt^2+a^2(t)\left[\frac{dr^2}{1-kr^2/R_0^2}+r^2\left(d\theta^2+\sin^2\theta
d\phi^2\right)\right],
\end{equation}
where $k=+1,0,-1$ corresponds to a closed, a flat,
and an open universe, respectively, and $R_0$ is the radius of spatial curvature
of the universe in the present epoch. Assuming that the matter
contribution is in the form of an ideal fluid with energy
density $\rho$ and pressure $p$, Park obtained \cite{ParkJHEP09}
\begin{eqnarray}
\left(\f{\dot{a}}{a}\right)^2&=&\frac{\kappa^2}{6(3\lambda-1)}
\left[\rho \pm \frac{3\kappa^2\mu^2}{8(3\lambda-1)} \left( \f{-
k^2}{R_0^4 a^4}+ \f{2 k (\La_W -\om)}{R_0^2 a^2}- \La_W^2 \right) \right] ,
\label{Friedmann Eq} \\
\f{\ddot{a}}{a}&=&\frac{\kappa^2}{6(3\lambda-1)} \left[-\f{1}{2}
(\rho+3 p) \pm \frac{3 \kappa^2\mu^2}{8(3\lambda-1)} \left( \f{
k^2}{R_0^4 a^4}- \La_W^2 \right) \right],
\end{eqnarray}
where the analytic continuation $\mu^2 \rightarrow - \mu^2$ for
$\Lambda_W$ has been employed \cite{ LuPRL09,
ParkJHEP09,ParkCQG08}. The upper (lower) sign corresponds to the
case where  $\La_W<0$ ($\La_W>0$).

Comparing them with the Einstein equations derived from GR with
the FRW metric,
\begin{eqnarray}
\left(\f{\dot{a}}{a}\right)^2&=&\frac{8 \pi G }{3 c^2}
(\rho_{\rm m} +\rho_\de )-\f{c^2  {k} }{R_0^2 a^2},  \\
\f{\ddot{a}}{a}&=&-\frac{4 \pi G }{3 c^2} \left[(\rho_{\rm m} +
\rho_\de)+ 3 (p_{\rm m} + p_\de) \right] ,
\end{eqnarray}
one can connect the Ho\v{r}ava parameters $\kappa,\lambda,\mu,
\Lambda_W,\omega$  with the speed of light $c$, the Newton's
constant $G$, and the effective dark energy density $\rho_\de$ and
pressure $p_\de$, although the connection is not unique. Park
defined the fundamental constants $c$ and $G$ as
\begin{equation}
c^2 = \f{ \kappa^4 \mu^2 |\La_W|}{ 8 (3 \la-1)^2},~G=\f{\kappa^2
c^2}{16 \pi (3 \la -1)} . \label{fundamental}
\end{equation}
To ensure the positivity of the dark energy density as required by
observations, we consider the case where $\Lambda_W>0$ and
identified the dark energy density and pressure as
\begin{\eq}
\rho_\de&=&\frac{3 c^4}{16\pi G \Lambda_W}\left( \frac{H_0^4 \Omega_k^2}{c^4 a^4}-\frac{2 H^2_0 \omega \Omega_k}{c^2 a^2} + \Lambda_W^2\right), \label{rho} \\
p_\de&=&\frac{3c^4}{16\pi G \Lambda_W}\left(\frac{H^4_0 \Omega^2_k}{3 c^4 a^4}+\frac{2  H^2_0 \omega \Omega_k}{3 c^2 a^2} - \Lambda^2_W \right) ,   \label{pressure}
\end{\eq}
where $\Omega_k = -(c^2 k)/(R^2_0 H^2_0)$. The equation of state
parameter is then
\begin{\eq}
w_\de \equiv \f{p_\de}{\rho_\de}=\f{H^4_0 \Omega^2_k + 2 c^2 H^2_0
\omega \Omega_k a^2 - 3c^4\Lambda^2_W a^4 } {3H^4_0 \Omega^2_k - 6
c^2 H^2_0 \omega \Omega_k a^2 + 3 c^4 \Lambda^2_W a^4 }
\label{EOS} .
\end{\eq}
From Eq.\ (\ref{rho}) we obtain
\begin{eqnarray}
\frac{\rho_\de}{\rho_c}  & = & \displaystyle \left( \frac{H^2_0 \Omega^2_k}{2c^2 \Lambda_W}\right)\frac{1}{a^4}
                   - \left(\frac{\Omega_k \omega }{\Lambda_W} \right) \frac{1}{a^2}
                   + \frac{c^2}{2 H^2_0}\Lambda_W \label{Omega DE} \\
             & \equiv &\displaystyle \Omega_1 a^{-4} + \Omega_2 a^{-2} + \Omega_3, \label{Omega DE 2}
\end{eqnarray}
where the critical energy density $\rho_c = (3H^2_0c^2)/(8 \pi G)$ and
\begin{equation} \label{Omega123}
\Omega_1 = \frac{H^2_0 \Omega^2_k}{2c^2 \Lambda_W} \geq 0 , \quad %
\Omega_2 = - \frac{\Omega_k \omega}{\Lambda_W}, \quad %
\Omega_3 = \frac{c^2 \Lambda_W}{2 H^2_0} > 0 .
\end{equation}

As shown by the above formulae, this effective dark energy
consists of three components that are radiation-like,
curvature-like and CC-like, respectively. It involves three model
parameters: $\{\Omega_k,\omega, \Lambda_W \}$. The first parameter
stems from the FRW metric ansatz, and the last two from the
Ho\v{r}ava gravity action. Note that for a flat universe this
effective dark energy behaves as a CC and accordingly this model
is the same as the CC dark energy model in GR, i.e.\ flat
$\Lambda$CDM, which is consistent with all the current
observational results. In the present paper we will consider a
nonzero $\Omega_k$ for possible deviation from $\Lambda$CDM.

\section{Model Parameters and Phenomenological Parameters}


To compare the Ho\v{r}ava Effective Dark Energy (\emph{HEDE})
model with observational results, one effective approach is to
employ a phenomenological parametrization of the relevant
quantities, whose observational constraints have been well studied,
as a mediator to facilitate the comparison. The observational
constraints on the model can be obtained from those on the
phenomenologically parametrized quantities by invoking an approximate
relation between the model parameters and the phenomenological parameters.

In \emph{HEDE} the model parameters include $\{ \Omega_k , \omega
, \Lambda_W \}$, which determine the dark energy behavior. That
is, the dark energy density $\rho_\de (a)$, as well as $w_\de
(a)$, is a function of $\{ \Omega_k , \omega , \Lambda_W \}$.
Phenomenologically, the evolution of dark energy is determined by
the present density fraction $\currentDE$ and its equation of
state parameter $w_\de$.  Here we invoke the widely used CPL
parameterization \cite{Chevallier00IJMP, LinderPRL03} of $w_\de$
in Eq.\ (\ref{CPL}): $w_\de = w_0 + w_a (1-a)$. In addition, since
the spatial curvature $\Omega_k$ is involved in the model
parameters, it should also be included in the phenomenological
parameter space when connecting to the model space. In summary,
the phenomenological parameters are $\{ \Omega_k , \currentDE
, w_0 , w_a \}$. Accordingly, we have a three-dimensional model
parameter space and a four-dimensional phenomenological parameter
space. A mapping between them is required for constraining the
model via the observational constraints on the phenomenological
parameters.

The relation between $\currentDE$ and $\{ \Omega_k , \omega ,
\Lambda_W \}$ can be obtained from Eq.\ (\ref{Omega DE}) by
setting $a=1$:
\begin{equation} \label{OmegaDE0}
\currentDE (\Omega_k,\omega,\Lambda_W) =
\frac{H^2_0 \Omega^2_k}{2c^2 \Lambda_W}
- \frac{\Omega_k \omega }{\Lambda_W}
+ \frac{c^2}{2 H^2_0}\Lambda_W .
\end{equation}
Following Park \cite{ParkJCAP10}, we connect the model parameters
$\{ \omega , \Lambda_W \}$ with the phenomenological parameters by
firstly expanding $w_\de$ in (\ref{EOS}) around $a=1$ as
\begin{equation}
w_\de = \left. w_\de \right|_{a=1} - \left. w'_\de \right|_{a=1}(1-a) + \cdots,
\end{equation}
where the prime denotes the derivative with respect to $a$, and
then identify $w_\de|_{a=1}$ and $-w'_\de|_{a=1}$ as $w_0$ and
$w_a$ in the CPL parameterization. As a result, the approximate
relation between $\{\omega, \Lambda_W\}$ and $\{w_0, w_a\}$ reads
\begin{eqnarray}
w_0 (\Omega_k,\omega,\Lambda_W) &=&
\frac{H_0^4 \Omega_k^2 + 2c^2H_0^2\omega \Omega_k - 3c^4\Lambda_W^2}
{3H_0^4 \Omega_k^2 - 6c^2H_0^2\omega \Omega_k + 3c^4\Lambda_W^2} ,
\label{w0} \\
w_a (\Omega_k,\omega,\Lambda_W) &=&
- \frac{8 c^2 H_0^2 \Omega_k \left( H_0^4\omega \Omega_k^2 - 2c^2H_0^2\Omega_k \Lambda_W^2 + c^4\omega \Lambda_W^2 \right)}
{3 \left( H_0^4 \Omega_k^2 - 2c^2H_0^2\omega \Omega_k + c^4\Lambda_W^2 \right)^2} ,
\label{wa}
\end{eqnarray}
or equivalently,
\begin{eqnarray}
\om &=& -\f{(1-2 w_0-3 w_0^2 -w_a)H^2_0 \Omega_k}{(1+4 w_0+3 w_0^2 +w_a)c^2}, \label{omega} \\
\La_W^2 &=& \f{(-1+9 w_0^2 +3w_a) H^4_0 \Omega_k^2}{3(1+4 w_0+3 w_0^2 +w_a)c^4} . \label{Lambda}
\end{eqnarray}
The required mapping between  $\{ \Omega_k , \omega , \Lambda_W
\}$ and $\{ \Omega_k , \currentDE , w_0 , w_a \}$ is given by
Eqs.\ (\ref{OmegaDE0}), (\ref{w0}) and (\ref{wa}). It maps the 3D
model space to a 3D hypersurface in the 4D phenomenological
parameter space.

Regarding the observational constraints, as stated in Sec.\ 1, we
assume $\currentDE \approx 0.74$, $\Omega_k<0.01$
\cite{7yearWMAP1}, $\Omega_1 < \Omega_r$, and the constraint on
$w_\de$ obtained in \cite{ChenPLB09}. For simplicity, we fix
$\currentDE=0.74$. This reduces the dimension of the
phenomenological parameter space from four to three: $\{\Omega_k,
w_0, w_a \}$, and that of the model space from three to two, i.e.\
a 2-dimensional surface in the 3D space $\{ \Omega_k , \omega ,
\Lambda_W \}$. This 2D surface corresponds to the relation
$\Omega_k = \Omega_k (\omega , \Lambda_W)$ obtained from Eq.\
(\ref{Omega DE}) with $\rho_\de(a=1)/\rho_c = 0.74$. In this case
the relations in Eqs.\ (\ref{OmegaDE0}), (\ref{w0}) and (\ref{wa})
become $\currentDE = \currentDE (\omega,\Lambda_W)$, $w_0 = w_0
(\omega,\Lambda_W)$ and $w_a = w_a (\omega,\Lambda_W)$. They map
the reduced 2D model space $\{ \Omega_k(\omega ,
\Lambda_W),\omega,\Lambda_W \}$ to a 2D surface $\{\Omega_k(\omega
, \Lambda_W), w_0(\omega , \Lambda_W), w_a(\omega , \Lambda_W) \}$
in the reduced 3D phenomenological parameter space $\{\Omega_k,
w_0, w_a \}$. Then, the observational constraint on the model is
represented by the intersection of this 2D surface and the well-studied,
observationally allowed regions of the 3D
phenomenological parameter space. We note that the spatial
curvature is well constrained. The upper bound of $|\Omega_k|$ is
around $0.01$ and could be even smaller \cite{7yearWMAP1}. Thus,
the allowed region in the 3D phenomenological parameter space is
very thin in the $\Omega_k$ direction. Accordingly, as a good
approximation after imposing the constraint $\Omega_k < 0.01$, we
will simply consider the constraints on the $\{ w_0,w_a \}$ plane
when taking the above-mentioned intersection that presents the
valid region of the model.

\begin{figure}[H]
\centering
\includegraphics[height=0.3\textheight]{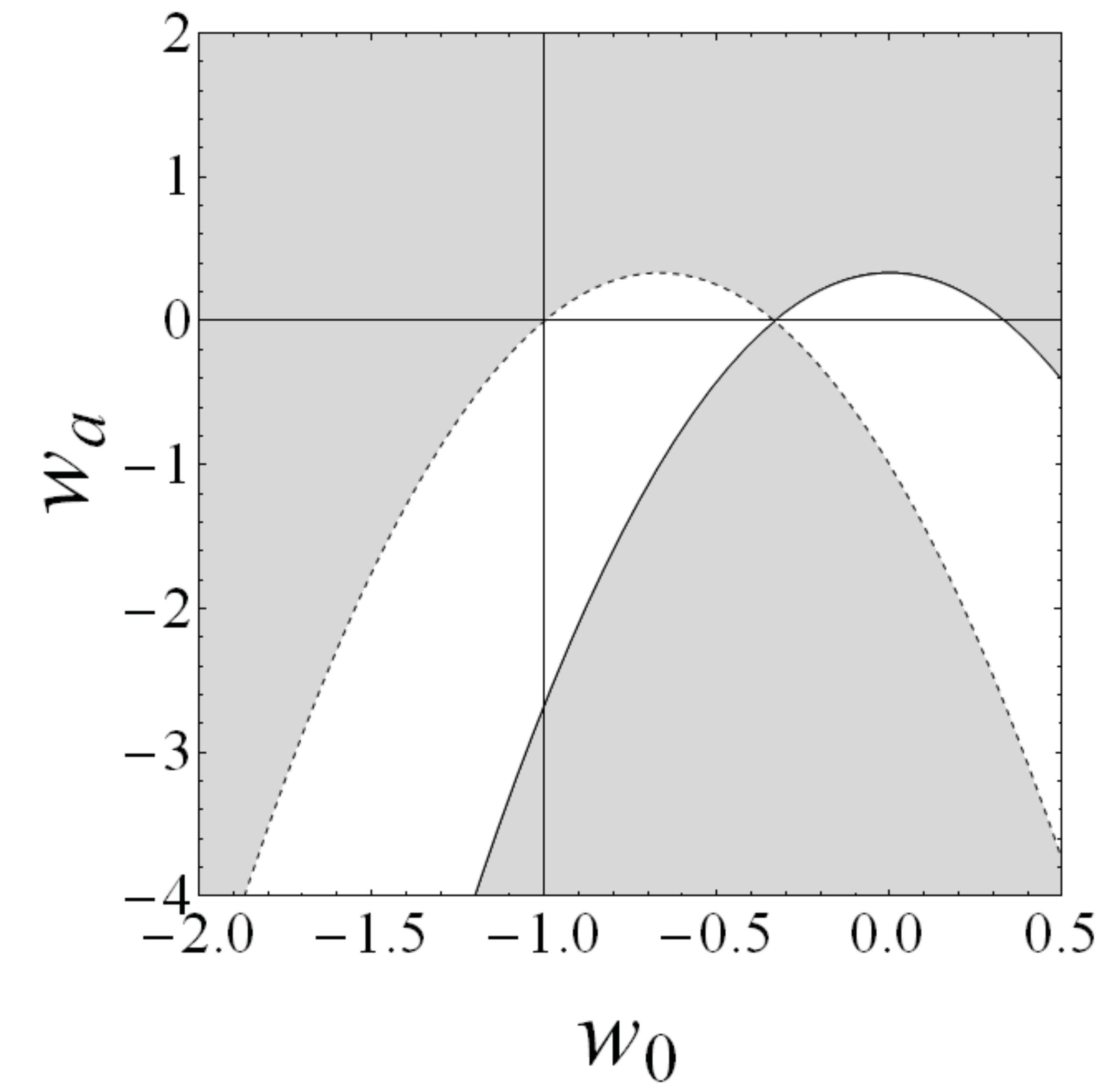}
  \caption{The valid (shaded) region for the self-consistency condition $\Lambda_W^2 > 0.$} \label{constraint1plot}
\end{figure}

\section{Constraints on the Phenomenological Parameter Space}

x
As commented in \cite{ParkJCAP10}, for the sake of self-consistency one should require
$\Lambda_W^2\geq 0$. From Eq.\ (\ref{Lambda}), this in turn requires that
\begin{eqnarray}
&& \left\{ w_a > -1-4 w_0-3 w_0^2, ~ w_a \geq (1-9 w_0^2)/3 \right\} \nonumber \\
&\mbox{or }&
\left\{ w_a<-1-4 w_0-3 w_0^2, ~ w_a \leq (1-9 w_0^2)/3 \right\} . \label{constraint0}
\end{eqnarray}
To simplify the following calculations, we define
\begin{eqnarray}
A & = & 1+ 4w_0 + 3w_0^2 +w_a, \label{A} \\
B & = & -1 + 9w^2_0 + 3w_a.    \label{B}
\end{eqnarray}
The self-consistency condition in Eq.\ (\ref{constraint0}) then reads
\begin{equation}
\{ A>0 , B \geq 0 \} \quad \mbox{or} \quad \{ A<0 , B \leq 0 \} . \label{constraint00}
\end{equation}
The valid region on the $\{ w_0 , w_a \}$ plane for this
self-consistency condition is presented in Figure
\ref{constraint1plot} by the shaded area.

Rewriting Eqs.\ (\ref{omega}) and (\ref{Lambda}) in terms of $A$
and $B$, we have
\begin{equation}
\omega = \frac{3A+B-8}{6A}\left( \frac{H^2_0 \Omega_k}{c^2}\right), ~ \Lambda_W = \sqrt{\frac{B}{3 A}}\left( \frac{H^2_0}{c^2}\right)|\Omega_k|. \label{om lm}
\end{equation}
Substituting Eq.\ (\ref{om lm}) into Eq.\ (\ref{OmegaDE0}), we obtain
\begin{equation}
|\Omega_k|= \mbox{sgn}(A) \frac{\sqrt{3}}{4}\sqrt{AB} ~ \currentDE ,
\label{Omega k}
\end{equation}
where sgn$(A)$ denotes the sign of $A$. For a positive dark energy density, this relation requires $A \geq 0$, which, as combined with Eq.\  (\ref{constraint00}), leads to
\begin{equation}
\{ A>0 , B\geq 0 \} , \label{constraint000}
\end{equation}
thereby excluding the bottom middle shaded area in Figure \ref{constraint1plot}.

\begin{figure}[H]
\centering
\includegraphics[height=0.3\textheight]{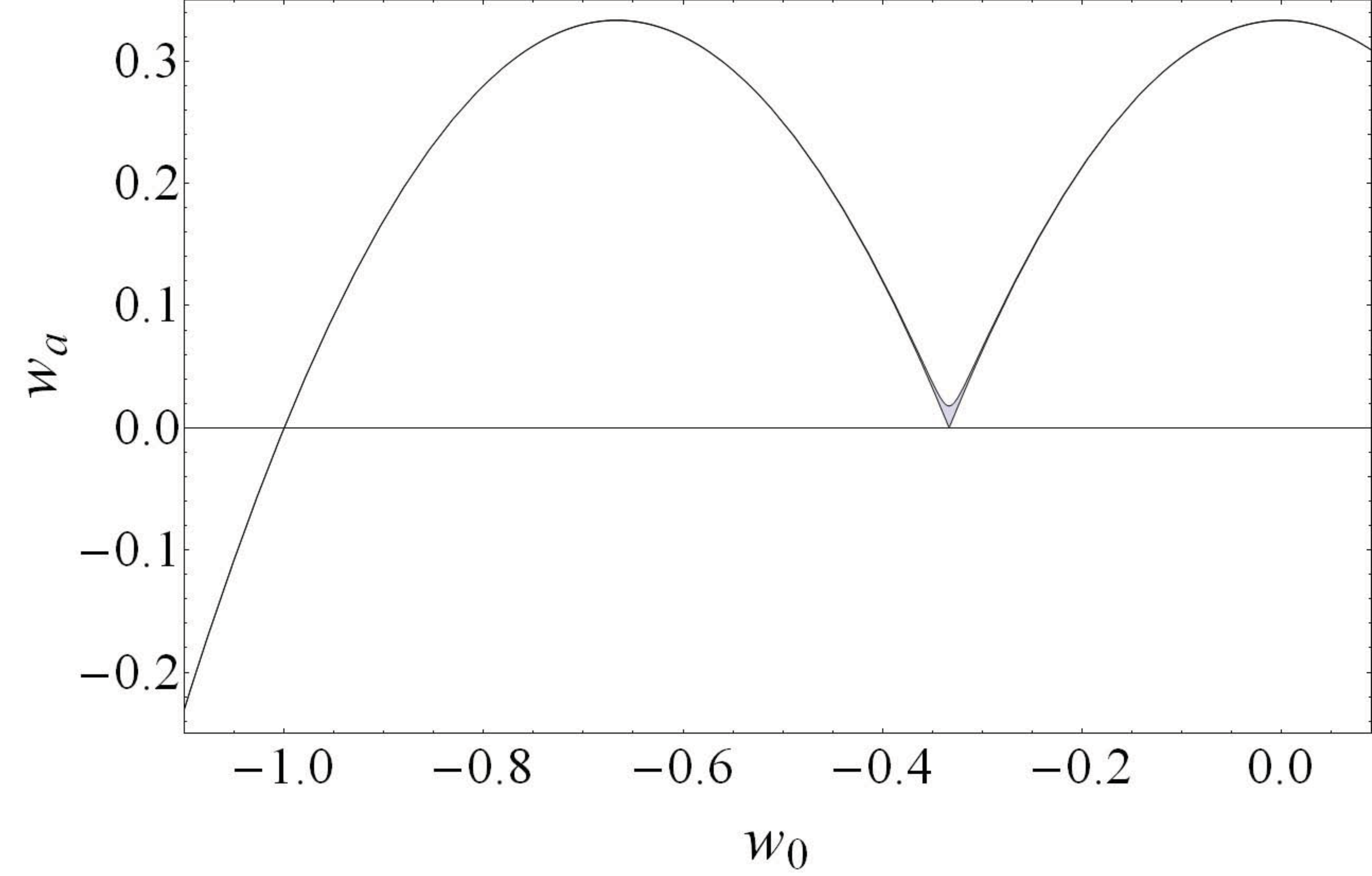}
\caption{The shaded area shows the largely reduced allowed region
for the observational requirements,
$\currentDE \approx 0.74$ and $|\Omega_k|<0.01$,
that lead to the constraint in Eq.\ (4.8).}
\label{constraint2plot}
\end{figure}

The current observations
suggest $\currentDE \approx 0.74$ and $|\Omega_k|<0.01$
\cite{7yearWMAP1}. With Eqs.\ (\ref{Omega k}) and (\ref{constraint000}), these two
requirements give a stringent constraint on the parameters $A$ and $B$
(i.e.\ $w_0$ and $w_a$):
\begin{equation}
0< AB < \left( \frac{4}{\sqrt{3}} \frac{0.01}{0.74} \right)^2 \cong 9.74\times 10^{-4}.
\label{constraint1}
\end{equation}
This constraint largely shrinks the allowed region in the
parameter space, which is presented by the shaded region in Figure
\ref{constraint2plot}.

As indicated in Eq.\ (\ref{Omega DE 2}), the effective dark energy
sector of the (IR-modified) \emph{HEDE} model consists of a
radiation-like ($\Omega_1$), a curvature-like ($\Omega_2$), and a
constant ($\Omega_3$) term. In the early universe when $a$ is
small, the radiation-like term dominates the effective dark energy
sector. Such a term, if too large, would cause the epoch of the
matter-radiation equality happened at a time that is later than that suggested by the
observational results about CMB and BBN.
Specifically, this term may contribute to the effective
relativistic degrees of freedom in the early universe. In the CMB
analysis the radiation is usually subdivided into two
categories: (i) photons and (ii) effective neutrinos (including
neutrinos and other effective relativistic particles).
Accordingly,
\begin{equation}
\Omega_r = \Omega_\gamma + \Omega_\nu = \Omega_\gamma (1+ 0.2271 N_{eff}),
\end{equation}
where $\Omega_r$ is the present radiation energy density fraction,
and $N_{eff}$ is the number of effective neutrino species. The
WMAP results suggest that $\Omega_r \cong 8.47 \times 10^{-5}$ and
$N_{eff} \approx 4$ \cite{7yearWMAP1}. Accordingly the effective
neutrinos have a comparable contribution to the radiation energy
density. Regarding the radiation-like term from Ho\v{r}ava gravity
as a source of the effective neutrinos, we obtain an upper bound
of $\Omega_1$ given by $0.2271 N_{eff}\Omega_\gamma$, which is of
the same order as $\Omega_r$.
This leads us to impose the constraint,
\begin{equation}
\Omega_1 < \Omega_r \cong 8.47 \times 10^{-5} . \label{constraint2}
\end{equation}

\begin{figure}[H]
\centering
\includegraphics[height=0.5\textheight]{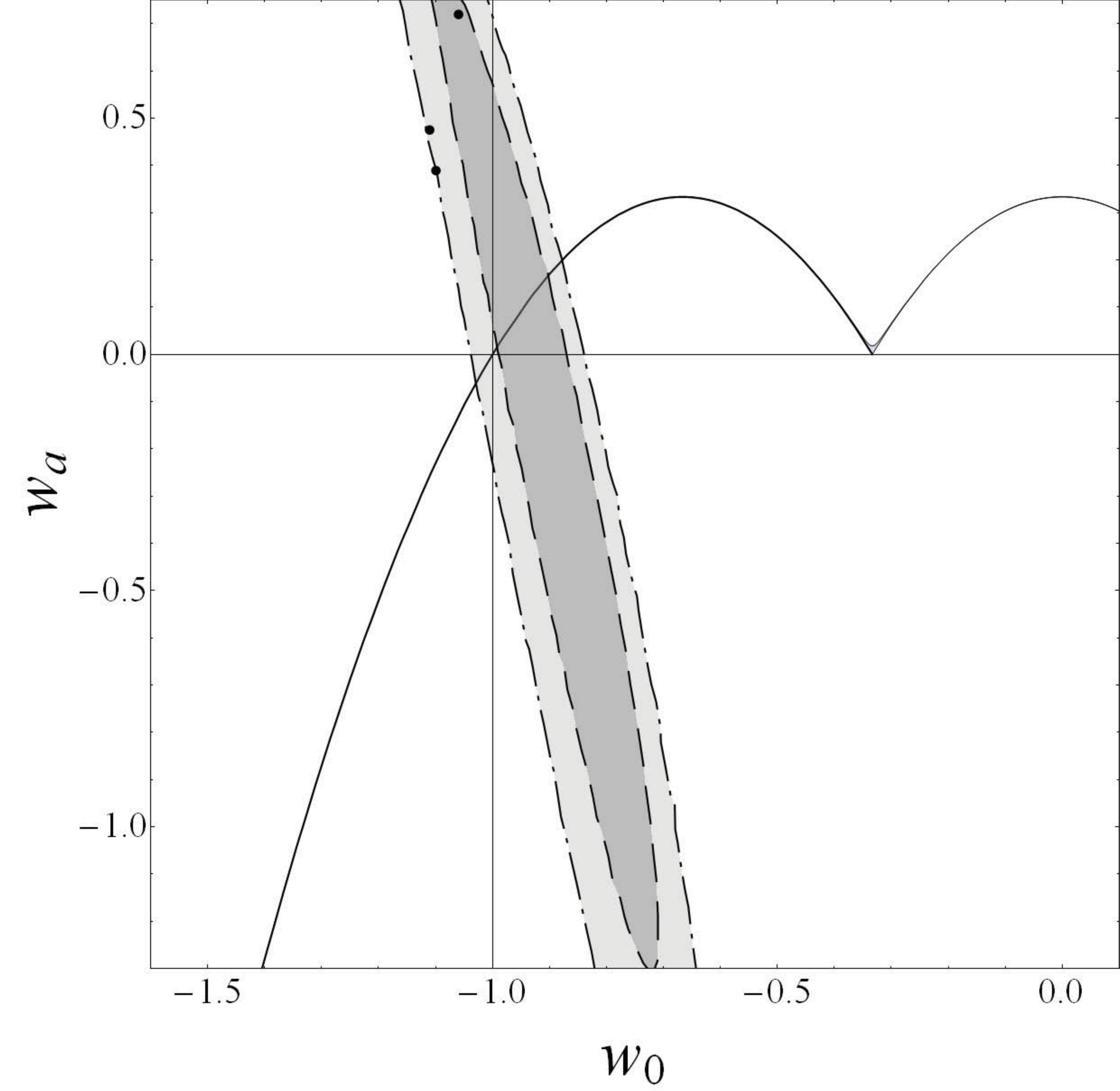}
\caption{Various constraints on the $w_0$-$w_a$ plane
for the IR-modified Ho\v{r}ava Effective Dark Energy (\emph{HEDE}) model.
The constraints are given in Eqs.\ (4.8) and (4.11),
and in \cite{ChenPLB09}. The long-dashed and the dot-dashed contour respectively
present the 1$\sigma$ and the 2$\sigma$ constraint on $\{ w_0 , w_a \}$,
which is obtained in \cite{ChenPLB09} from the current SN Ia, CMB and BAO data.
The narrow gray region presents the constraint in Eq.\ (4.8) for the requirement $|\Omega_k|<0.01$. The narrow black region that nearly overlaps with the narrow gray region for $w_0 < -1/3$ presents the constraint in Eq.\ (4.11) for the requirement $\Omega_1 < \Omega_r$.
The intersection of these three kinds of allowed regions gives the valid \emph{HEDE} model.
The three black dots denote the best-fit models Park considered in \cite{ParkJCAP10}, none of which is in the valid region.}
\label{constraint3plot}
\end{figure}

Substituting Eqs.\ (\ref{om lm}) and (\ref{Omega k}) into the
definition of $\Omega_1$ in Eq.\ (\ref{Omega123}), we obtain
$\Omega_1 = 3 |A| \currentDE / 8$, from which and Eq.\ (\ref{constraint000}) the above constraint requires
\begin{equation}
0 < A < \frac{8 \Omega_r}{3 \currentDE} \approx 3.05\times 10^{-4}. \label{constraint21}
\end{equation}
This tightly constrained region is presented in Figure
\ref{constraint3plot} by the black area which is so narrow that it
looks like a black curve. In this figure we also show the
constraint given in Eq.\ (\ref{constraint1}), which is presented by the gray
area. For $w_0 < -1/3$, it largely overlaps with the black narrow region.

In addition, in Figure \ref{constraint3plot} we show the 1$\sigma$
(long-dashed contour) and the 2$\sigma$ (dot-dashed contour)
observational constraints of $w_0$ and $w_a$ obtained in
\cite{ChenPLB09} from the combined data set that includes the
SN-Ia data from the Constitution Set, the CMB measurement from the
five-year WMAP, and the BAO measurement from SDSS and 2dFGRS. The
intersection of all the above-mentioned allowed regions gives the
valid IR-modified Ho\v{r}ava Effective Dark Energy model, which is
the black narrow region enclosed by the long-dashed ($1\sigma$) or
the dot-dashed ($2\sigma$) contour.

\section{Constraining the IR-modified Ho\v{r}ava Gravity}

To constrain the IR-modified Ho\v{r}ava gravity, here we transfer the observational constraints on the phenomenological parameters to the \emph{HEDE} model via the mapping between the phenomenological parameter space and the model space. This mapping is given in Eqs.\ (\ref{omega}) and (\ref{Lambda}), i.e., in Eq.\ (\ref{om lm}), where $|\Omega_k|$ is a function of $A$ and $B$, as given in Eq.\ (\ref{Omega k}), after we fix $\currentDE=0.74$. With Eqs.\ (\ref{Omega k}) and (\ref{constraint000}) substituted into Eq.\ (\ref{om lm}), this mapping can be rewritten as
\begin{eqnarray}
\tilde{\omega} &=& s_k \frac{\sqrt{3}}{24} (3A+B-8) \sqrt{\frac{B}{A}} ,
\label{w map 50} \\
\tilde{\Lambda}_W &=& \frac{1}{4} B , \label{LambdaW map 50}
\end{eqnarray}
where the two dimensionless parameters $\tilde{\omega}$ and $\tilde{\Lambda}_W$ are defined as
\begin{equation}
\tilde{\omega}=\frac{c^2 \omega}{H^2_0 \currentDE} , \quad
\tilde{\Lambda}_W=\frac{c^2 \Lambda_W}{H^2_0 \currentDE} ,
\label{dimless omega & LambdaW}
\end{equation}
and $s_k$ denotes the sign of $\Omega_k$.

\begin{figure}[H]
\centering
\includegraphics[height=0.4\textheight]{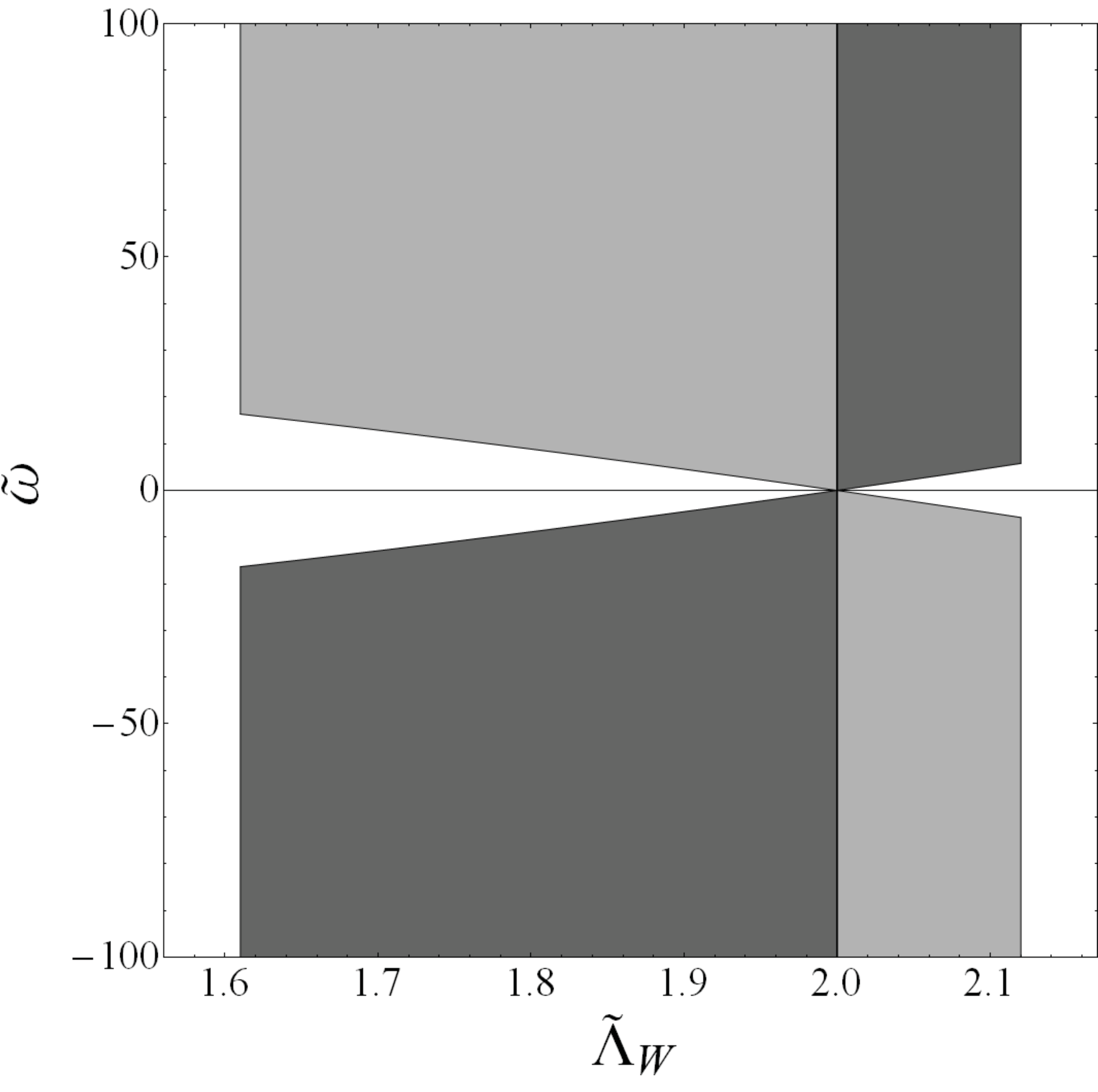}
\caption{The valid $\tilde{\omega}$-$\tilde{\Lambda}_W$ region corresponding to the valid region in the $w_0$-$w_a$ space. The dark and the light region correspond to $\Omega_k>0$ and $\Omega_k<0$, respectively. }\label{omegaLambdaWplot}
\end{figure}

In the valid region, i.e.\ the black narrow region enclosed by the $2\sigma$ contour in Figure \ref{constraint3plot}, we have the constraints $0<A<3.05\times 10^{-4}$, as required in Eq.\ (\ref{constraint2}), and $6.44<B<8.48$. From the above mapping and the constraints on $A$ and $B$, we obtain
\begin{equation}
\tilde{\omega} \left( \epsilon , \tilde{\Lambda}_W \right) =
\frac{s_k}{2}
\left[ \left( \tilde{\Lambda}_W - 2 \right) + \epsilon \right]
\sqrt{\frac{\tilde{\Lambda}_W}{\epsilon}},
\label{lambdaWomegarelation}
\end{equation}
where
\begin{equation}
0 < \epsilon \equiv \frac{3}{4}A < 2.29\times 10^{-4} ,
\end{equation}
\begin{equation}
-0.39 < \tilde{\Lambda}_W -2 < 0.12 .
\end{equation}
For more details, the constraint on $\{ \tilde{\omega} , \tilde{\Lambda}_W \}$ can be read as follows.
\begin{eqnarray}
-\infty < s_k \tilde{\omega} < s_k \tilde{\omega}(\epsilon_{max}) < 0
& \mbox{ as } & \tilde{\Lambda}_W \leq 2-\epsilon_{max} \, , \nonumber \\
-\infty < s_k \tilde{\omega} < s_k \tilde{\omega}(\epsilon_{max}) > 0
& \mbox{ as } & 2-\epsilon_{max} < \tilde{\Lambda}_W < 2 \, , \nonumber \\
0 < s_k \tilde{\omega} < \sqrt{\epsilon_{max}/2}
& \mbox{ as } &  \tilde{\Lambda}_W = 2 \, , \nonumber \\
\sqrt{\tilde{\Lambda}_W(\tilde{\Lambda}_W-2)} < s_k \tilde{\omega} < \infty
& \mbox{ as } & 2 < \tilde{\Lambda}_W \leq 2 + \epsilon_{max} \, , \nonumber \\
0 < s_k \tilde{\omega}(\epsilon_{max}) < s_k \tilde{\omega} <  \infty
& \mbox{ as } & \tilde{\Lambda}_W > 2+\epsilon_{max} \, ,
\label{model constraint}
\end{eqnarray}
where $\epsilon_{max} = 2.29\times 10^{-4}$. This constraint is presented in
Figure \ref{omegaLambdaWplot}, where the dark region and the light region correspond to $s_k = +$ and $s_k = -$, i.e.\ $\Omega_k>0$ and $\Omega_k<0$, respectively.

As shown in Figure \ref{omegaLambdaWplot}, $\tilde{\Lambda}_W$ is
restricted to values between $1.61$ and $2.12$, and $|\tilde{\omega}|$ has a
nontrivially lower limit but no upper limit. We note that for $\tilde{\omega}=0$
the allowed region is almost a point, specifically,
\begin{equation}
- 2.29\times 10^{-4}< \tilde{\Lambda}_W - 2 < 0 .
\end{equation}
That is, in the case where detailed balance is preserved ($\omega=0$) we need to fine-tune the value of $\Lambda$ for the \emph{HEDE} model to be consistent with observational results. Thus, the cosmological test strongly disfavors, although does not rule, the Ho\v{r}ava action that preserves detailed balance. On the contrary, for large $|\tilde{\omega}|$ the full range of $\tilde{\Lambda}_W$, $(1.61,2.12)$, is allowed. Moreover, there is no upper limit to $|\tilde{\omega}|$. Accordingly, the observational results suggest the breaking of the detailed balance condition.

Note that the curvature-like effective energy term in Eq.\ (\ref{Omega DE 2}), as originated from the soft violation of detailed balance, remains finite when $\tilde{\omega}$ goes to infinity. This can be seen in the following.
\begin{equation}
\Omega_2
= - \frac{\Omega_k \omega}{\Lambda_W}
= \frac{\currentDE}{8}(8-3A-B) ,
\end{equation}
where $\currentDE \approx 0.74$, $0<A<3.05\times 10^{-4}$ and $6.44<B<8.48$ for the valid region. We emphasize that even though at the action level the magnitude of the soft violation can be arbitrarily large with no upper limit, the corresponding effective energy may still be tightly constrained.

\section{Possible Behavior of the Effective Dark Energy}

Recall Eqs.\ (\ref{Omega DE}) and (\ref{Omega DE 2}), the \emph{HEDE} consists of the radiation-like ($\Omega_1 a^{-4}$), the curvature-like ($\Omega_2 a^{-2}$) and the constant-like ($\Omega_3$) sectors:
\begin{eqnarray}
\frac{\rho_\de}{\rho_c}  & = & \displaystyle \left( \frac{H^2_0 \Omega^2_k}{2c^2 \Lambda_W}\right)\frac{1}{a^4}
                   - \left(\frac{\Omega_k \omega }{\Lambda_W} \right) \frac{1}{a^2}
                   + \frac{c^2}{2 H^2_0}\Lambda_W  \\
             & \equiv &\displaystyle \Omega_1 a^{-4} + \Omega_2 a^{-2} + \Omega_3.
\end{eqnarray}
Substituting Eqs.\ (\ref{omega}), (\ref{Lambda}) and (\ref{Omega k}) into Eq.\ (\ref{Omega123}), we obtain
\begin{align}
\Omega_1 &= \frac{3\currentDE}{8}(1+ 4w_0 + 3w_0^2 +w_a) , \\
\Omega_2 &= \frac{-3\currentDE}{4} (-1+ 2 w_0 + 3 w_0^2 + w_a) , \\
\Omega_3 &= \frac{\currentDE}{8}(-1+9w^2_0 + 3w_a) .
\end{align}
Note that $\Omega_1 + \Omega_2 +\Omega_3 = \currentDE$ as required. The requirement $\currentDE = 0.74$ and the constraint on $\{w_0, w_a \}$ give an allowed region in the $\{\Omega_1, \Omega_2, \Omega_3\}$ space: a plane
$\Omega_1+\Omega_2+\Omega_3 = 0.74 $ bounded by the box
$\{ 0 < \Omega_1 < 8 \times 10^{-6} ,
-0.05 < \Omega_2 < 0.15 ,
0.59 < \Omega_3 < 0.78 \}$ corresponding to the $2\sigma$ contour on
$\{w_0, w_a \}$.

\begin{figure}[h]
\centering
\includegraphics[height=0.3\textheight]{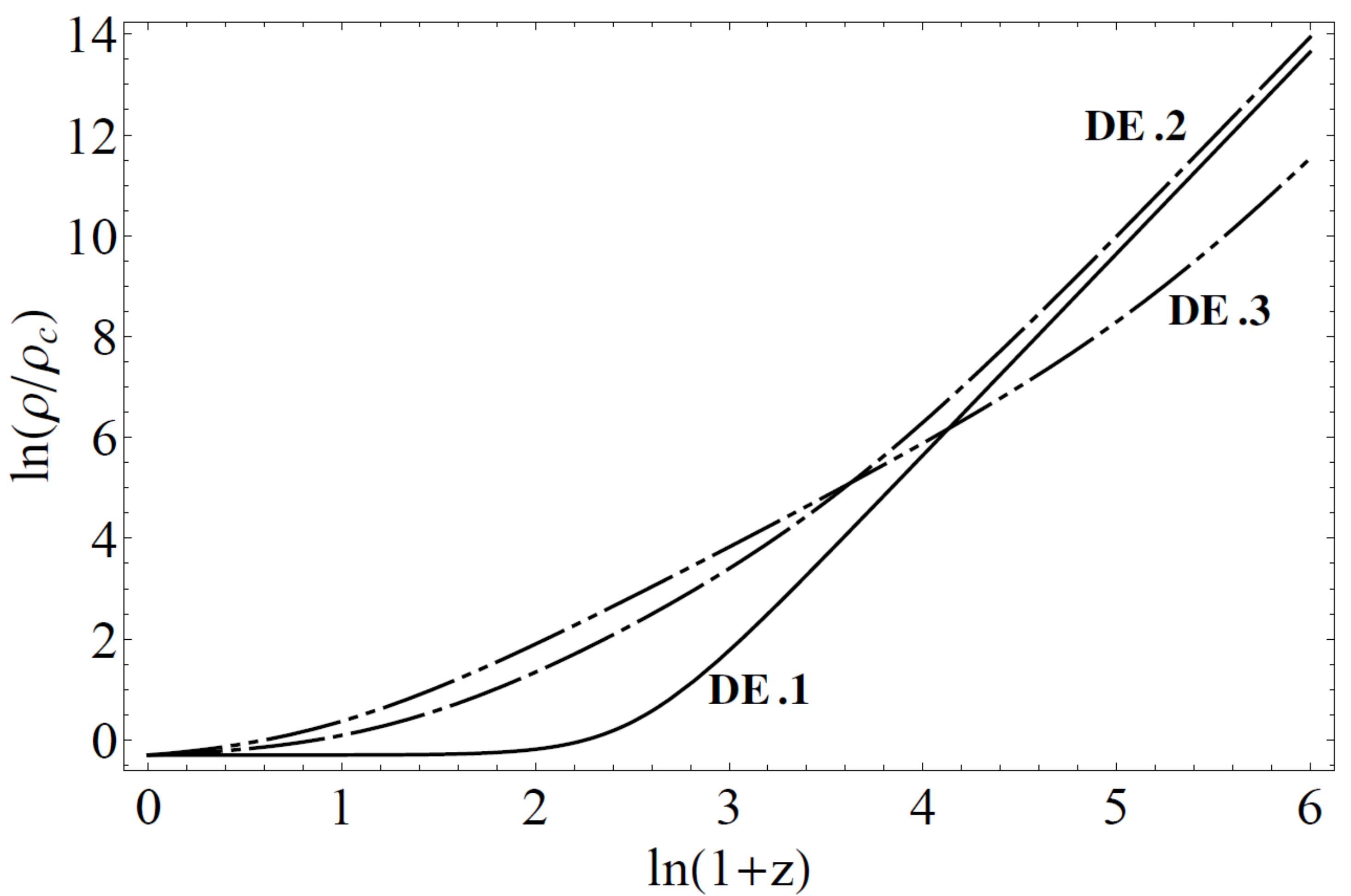}
  \caption{The energy density evolution patterns in three sample cases. The solid, the dot-dashed and the double-dot-dashed line correspond to $\mathbf{DE.1}$, $\mathbf{DE.2}$ and $\mathbf{DE.3}$, respectively.}\label{DEevolutionplot}
\end{figure}

To show the possible evolution patterns of the \emph{HEDE}, we consider three sample cases, $\mathbf{DE.1}$, $\mathbf{DE.2}$ and $\mathbf{DE.3}$, corresponding to three points in the narrow valid region in Figure  \ref{constraint3plot}. The values of $\{ w_0, w_a ; \Omega_1, \Omega_2, \Omega_3 \}$ in these three cases are as follows.
\begin{equation}
\begin{array}{c|cc|ccc}
 & w_0 & w_a & \Omega_1 & \Omega_2 & \Omega_3 \\
\hline
\mathbf{DE.1}~ & -1.00~ & 1.14\times 10^{-4} &
3.18\times10^{-5} & -6.35\times10^{-5} & ~0.740 \\
\mathbf{DE.2}~ & -0.95~ & 9.27\times10^{-2} &
4.24\times10^{-5} & ~~5.54\times10^{-2} & ~0.685 \\
\mathbf{DE.3}~ & -0.90~ & 1.70\times10^{-1} &
3.18\times10^{-6} & ~~1.11\times10^{-1} & ~0.629
\end{array}
\label{3DEcases}
\end{equation}
The evolution of the energy density in these three cases is shown in Figure \ref{DEevolutionplot}. These three cases have the same density fraction at present,  $\currentDE=0.74$, and share similar evolution patterns in the late times up to $\ln(1+z)\approx 0.25$. $\mathbf{DE.1}$ resembles $\rm \Lambda CDM$ at low redshifts. Its energy density remains nearly constant for $\ln(1+z) < 2.0$. This is because $\Omega_1$ and $\Omega_2$ are both extremely small in this case. In the cases of $\mathbf{DE.2}$ and $\mathbf{DE.3}$ the energy densities increase with $z$ rapidly for $\ln(1+z)> 0.25$ due to the $\Omega_2$ term. Generally speaking, it is the difference in $\Omega_2$ that makes the evolution patterns distinct from each other in the interval $0.25< \ln(1+z) < 3.5$. In the early universe, the radiation-like term in \emph{HEDE} would dominate. Hence for $\ln(1+z)>5$ the slopes of $\ln(\rho/\rho_c)$ versus $\ln(1+z)$ in different cases are roughly the same. The value of $\Omega_1$ determines the value of $\ln(\rho/\rho_c)$  at high redshifts. The cases with larger $\Omega_1$ have larger energy densities in the early times. Nevertheless, in the early times the energy density of \emph{HEDE} should be smaller than that of radiation, as required in Eq.\ (\ref{constraint2}).

\begin{figure}[h]
\centering
\includegraphics[width=0.5\textheight]{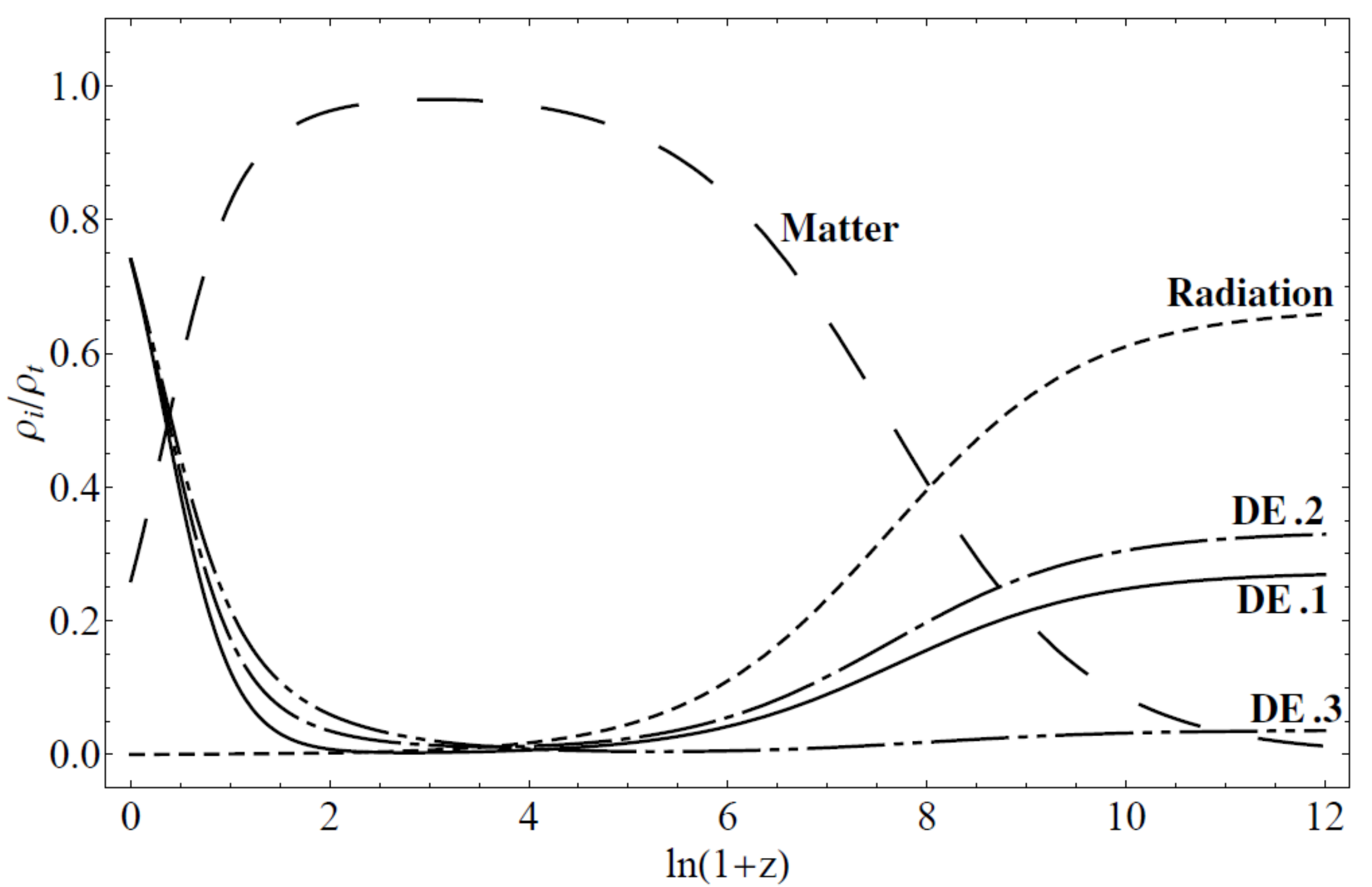}
\caption{The redshift dependence of the contribution of different energy contents to the total energy density. The solid, dot-dashed, double-dot-dashed, long-dashed, and short-dashed lines correspond to $\mathbf{DE.1}$, $\mathbf{DE.2}$, $\mathbf{DE.3}$, matter and radiation, respectively.}\label{Energyfraction}
\end{figure}

In Figure \ref{Energyfraction} we show the redshift dependence of the ratio $\rho_i/\rho_t$, where $\rho_i$ stands for the energy density of $\mathbf{DE.1}$, $\mathbf{DE.2}$, $\mathbf{DE.3}$, matter or radiation, and $\rho_t$ for the total energy density. The evolution of the three cases nearly coincide with each other after the time when  $\ln (1+z)\approx0.5$ (i.e.\ $z \approx 0.65$), around which the crossing between the \emph{HEDE} and the matter energy density at low redshift happens. This is because after the crossing the constant-like ($\Omega_3$) term dominates the dark sector, while these three cases have similar $\Omega_3$. Before this crossing, the three cases behave differently. For $\ln(1+z)<4$ the contribution of the \emph{HEDE} to the total energy decreases with $z$, and the decreasing is more rapid in $\mathbf{DE.1}$ than in $\mathbf{DE.2}$ and $\mathbf{DE.3}$. The main reason is that $\Omega_2$ in $\mathbf{DE.1}$ is negative. In general, $\rho_\de/\rho_t$ decreases with $z$ more rapidly for smaller $\Omega_2$. For $\ln(1+z)> 4$ (i.e., for $z>50$), the contribution of the \emph{HEDE} to the total energy increases with $z$. This is a reflection of the dominance of the radiation-like ($\Omega_1$) term over the other two terms in the dark sector. The slop of this increment and the contribution of the \emph{HEDE} at high redshifts are determined by the value of $\Omega_1$. In the case with larger $\Omega_1$, $\rho_\de$ increases with $z$ more rapidly and $\rho_\de / \rho_t$ is larger for $\ln(1+z)>4$. However, the contribution of the \emph{HEDE} would be smaller than that of radiation in the early times, as required in Eq.\ (\ref{constraint2}).

\begin{figure}[h]
\centering
\subfloat[$\mathbf{DE.1}$]{\includegraphics[width=0.48\textwidth]{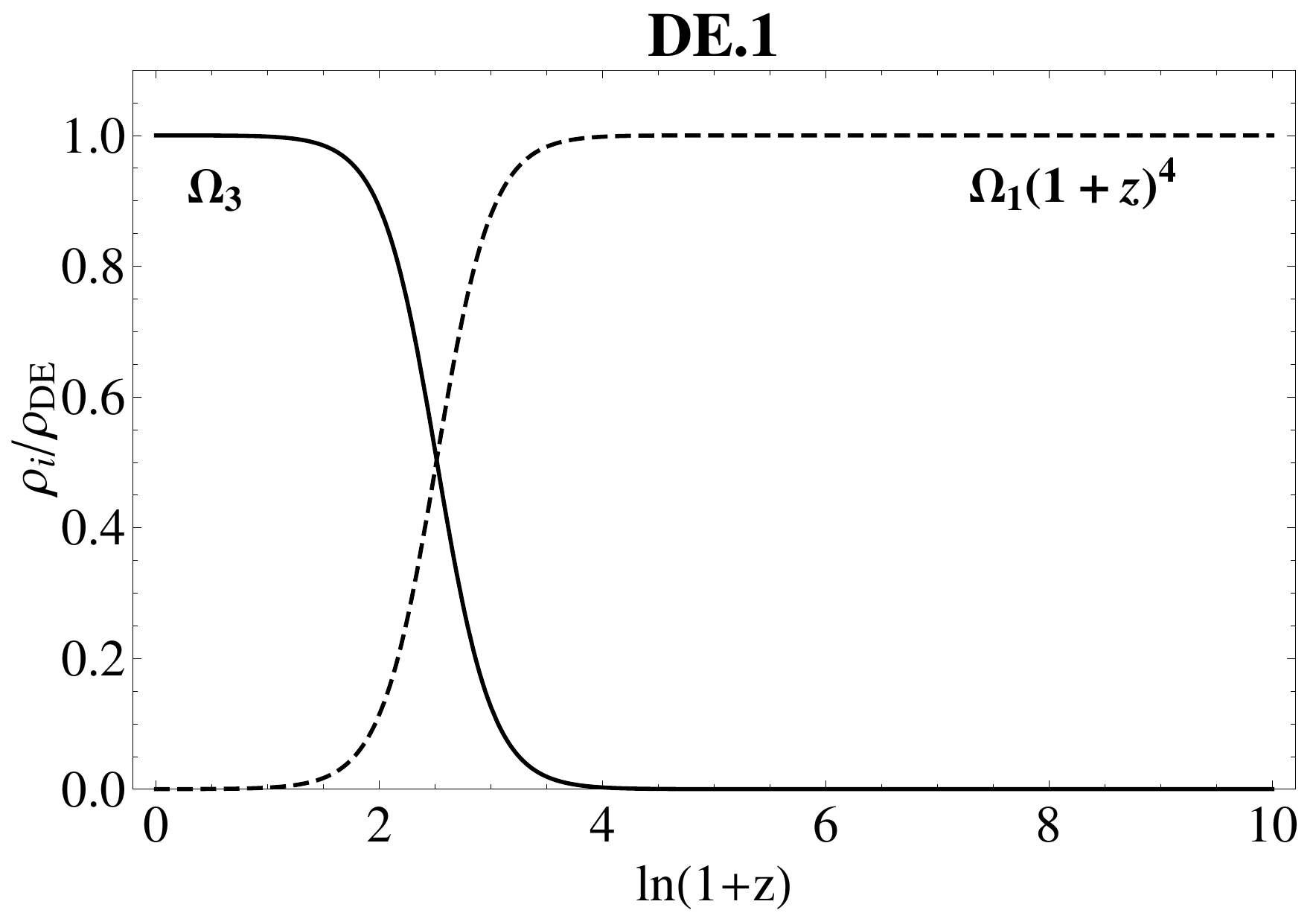}}\;
\subfloat[$\mathbf{DE.2}$]{\includegraphics[width=0.48\textwidth]{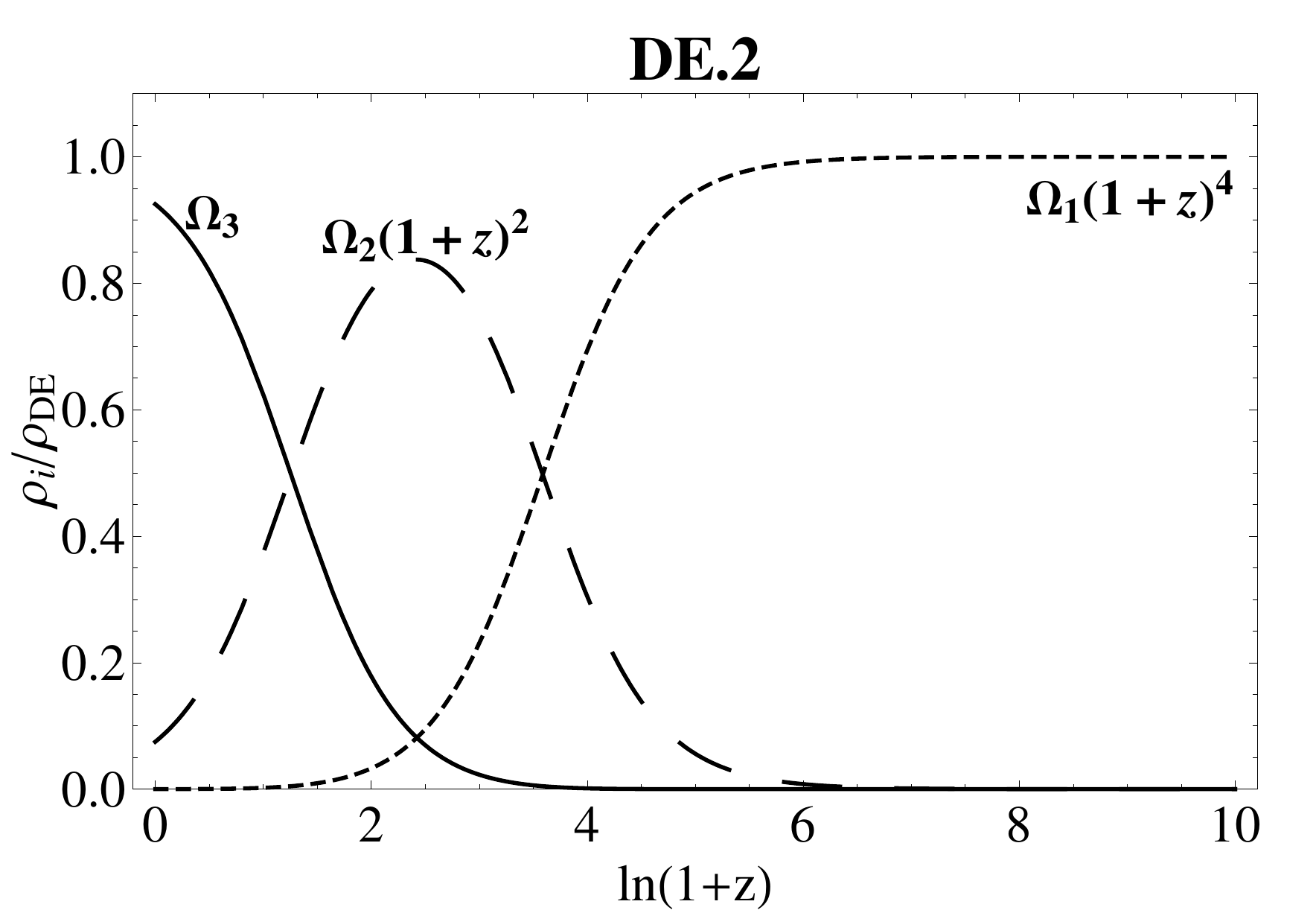}}\\
\subfloat[$\mathbf{DE.3}$]{\includegraphics[width=0.48\textwidth]{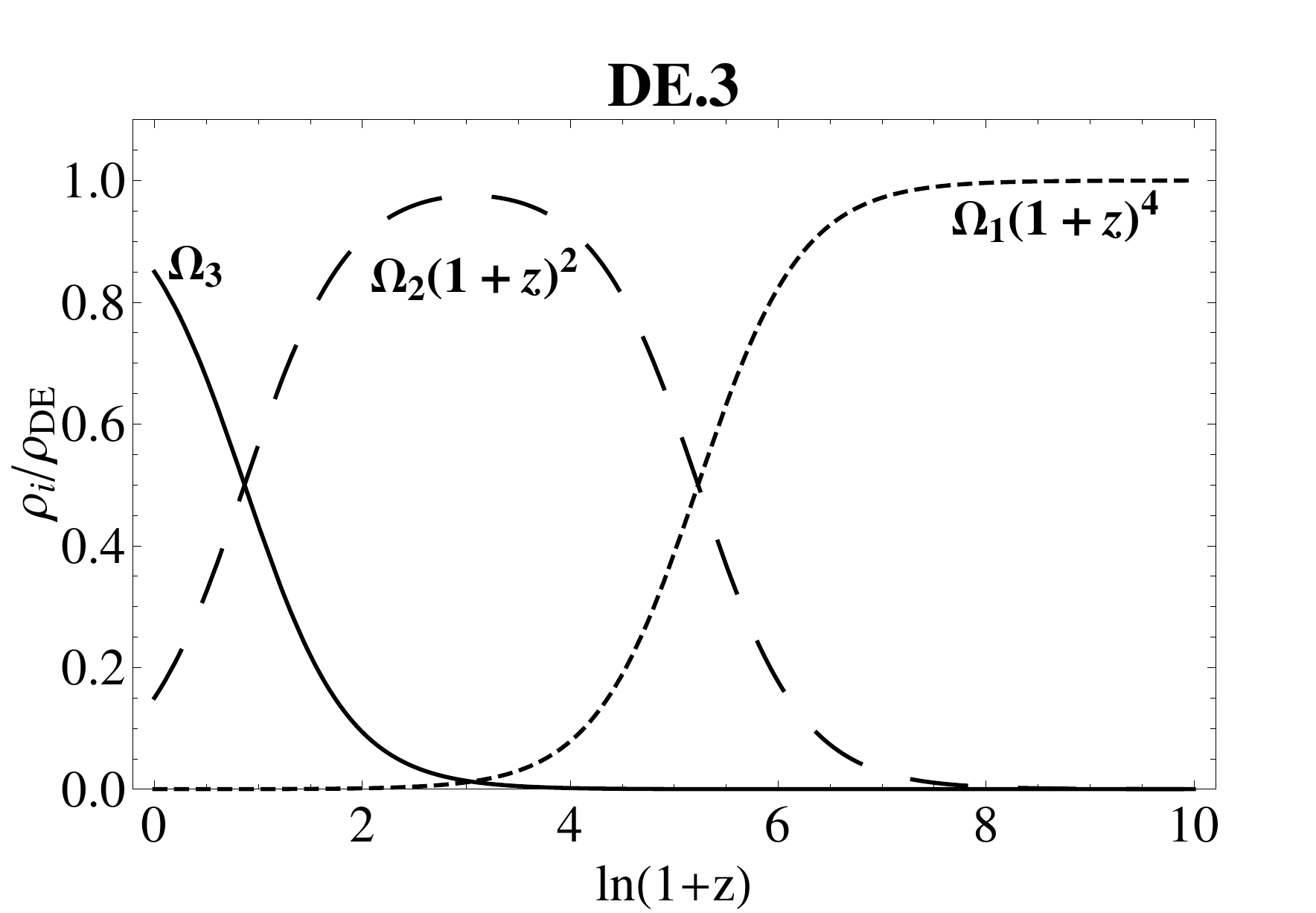}}
\caption{The redshift dependence of the contribution of the three dark terms in the \emph{HEDE} energy density for the three cases, $\mathbf{DE.1}$, $\mathbf{DE.2}$ and $\mathbf{DE.3}$. The solid, the long-dashed, and the short-dashed line correspond to the constant-like, the curvature-like, and the radiation-like term, respectively. }\label{DEfraction}
\end{figure}

In Figure \ref{DEfraction} we show the redshift dependence of the contribution of the three terms, $\{ \Omega_1 (1+z)^4, \Omega_2 (1+z)^2, \Omega_3 \}$, in the \emph{HEDE} energy density for the three cases in Eq.\ (\ref{3DEcases}). For all the three cases, the constant-like ($\Omega_3$) term dominates the dark sector in the late times, while the radiation-like ($\Omega_1$) term dominates in the early times. Whether there exists a period of the dominance of the curvature-like ($\Omega_2$) term depends on the magnitude of $\Omega_2$. For $\mathbf{DE.1}$, the contribution of the term $\Omega_2 (1+z)^2$ is much smaller than the other two terms. Hence in Figure \ref{DEfraction}a the line corresponding to this term cannot be seen. For the $\mathbf{DE.2}$ and $\mathbf{DE.3}$ cases, the curvature-like term dominates the dark sector during a ``middle age". The precise period of such ``middle age" varies from case to case. In general, for larger $\Omega_2$, the dominance of the curvature-like term starts earlier and ends later, with larger contribution from this term to the \emph{HEDE}.

\section{Summary and Discussion}

In this paper we test the IR-modified Ho\v{r}ava gravity from the
cosmological point of view, in particular, from the viewpoint of
cosmic expansion. We conclude that the Ho\v{r}ava
gravity with soft violation of the detailed balance condition is consistent with
the current observational results on the expansion history of
the universe. Specifically, this gravity theory can generate the
late-time cosmic acceleration with the behavior that is well consistent
with observations. We note that the Ho\v{r}ava gravity with
the detailed balance condition, though not ruled out, requires fine-tuning
$\Lambda_W$ such that $- 2.29\times 10^{-4}< (c^2 \Lambda_W)/(H^2_0 \currentDE) - 2 < 0 $
in order to fit the observational data. This result, together with
previous studies \cite{Calcagni:2009PRD, LuPRL09}, suggests that
the breaking of the detailed balance condition, at least softly, is
necessary to render Ho\v{r}ava gravity a more realistic IR-limit.

We obtained the observational constraints on two model parameters,
$\Lambda_W$ and $\omega$, i.e., the cosmological constant of the
three dimensional Einstein-Hilbert action and the coefficient of
the soft violation term. The parameter $\Lambda_W$ is well-constrained and
it should be of the order of the inverse square of
the Hubble length, $H_0^2/c^2$. More precisely, we found
that $\Lambda_W$, in units of $(H^2_0 / c^2)$, is bounded within a small
range, $(1.61\currentDE,2.12\currentDE)$, i.e., $(1.19,1.57)$
after imposing $\currentDE = 0.74$. On the other hand, we obtained
a lower bound, but without an upper bound, to $|\omega|$ with regard to the extent
of the soft violation of the detailed balance condition. The lower bound depends
on $\Lambda_W$, and in most cases it is also around the order of
the inverse square of the Hubble length, $H_0^2/c^2$.

With our more comprehensive investigation into the cosmic-expansion
test of Ho\v{r}ava gravity, we found the
Ho\v{r}ava effective dark energy (\emph{HEDE}) much more
restrictive than that deduced in \cite{ParkJCAP10}. Specifically, the
allowed parameter space $\{w_0, w_a\}$ is now much smaller. It is
a narrow strip beside the parabola $1+ 4w_0 + 3w_0^2 +w_a=0$
around $(w_0, w_a) = (-1, 0)$. The energy density of the \emph{HEDE}, with different values of the model parameters, can give rise to a cosmological
constant as well as non-constant behaviors. For the
latter cases with non-constant energy density, the main difference
between the models therein is the evolution of the dark energy density
in the ``middle age". This is because during the ``middle age" the curvature-like ($\Omega_2$) term dominates, while the range of allowed $\Omega_2$ is
not small, namely from $-0.04$ to $+1.03$. As a result the dark energy behavior
in the non-constant cases almost coincide with each other at high
redshifts. This is because in the early times the dark energy is
dominated by the radiation-like ($\Omega_1$) term that is highly restricted, namely
$\Omega_1\lesssim8.47\times10^{-5}$.

In our analysis we compare \emph{HEDE} with observational results
by invoking the CPL parametrization of the equation of state of
dark energy: $w_\de = w_0 + w_a (1-a)$, as a mediator.
We then transfer the constraint on the
phenomenological parameters $\{ \Omega_k, \currentDE, w_0, w_a \}$
to that on the model parameters $\{ \Omega_k, \omega, \Lambda_W
\}$ via an approximate relation between $\{ w_0, w_a \}$ and $\{
\Omega_k, \omega, \Lambda_W \}$. Naively, since the dimension of
the phenomenological parameter space is larger than that of the
model space, the 4D parameter space seems flexible enough to
accommodate the mapping from the 3D model space.
Nevertheless, to be prudent, the validity of this
approximation should be checked. In particular, it is important to verify that the two energy densities,
$\rho_\textrm{HEDE}(z;\Omega_k,\omega,\Lambda_W)$ and
$\rho_\textrm{CPL}(z;\currentDE,w_0, w_a)$, are consistent with each other,
where $\rho_\textrm{HEDE}$ is given in Eqs.\ (\ref{rho}) or (\ref{Omega DE}) and
\begin{equation}
\rho_\textrm{CPL}=\rho_0 e^{3w_a(a-1)}a^{-3(1+w_0+w_a)}.
\end{equation}
The approximation is valid when the difference between these
two energy densities is significantly smaller than the observational accuracy
in the relevant redshift range. This consistency check requires further investigation.
One way to avoid the possible incompatibility between the model space and the phenomenological parameter space is to use the model to fit data directly, e.g., invoking the $\chi^2$ fitting to obtain the observational constraints on the model
parameters $\{  \Omega_k, \Lambda_W, \omega \}$. This is under our investigation and will be reported in our follow-up paper.

Ho\v{r}ava gravity, an interesting alternative gravity theory that breaks the Lorentz symmetry, should ideally be constrained by observations and experiments ranging from microscopic, solar, astronomical, to cosmological scales. From the cosmological point of view, a modified gravity theory changes not only the cosmic expansion history but also the structure formation. In the present paper we have shown how the IR-modified Ho\v{r}ava gravity can be tightly constrained by the observations about the cosmic expansion. In addition to the expansion history, we expect the observations about the cosmic structures, such as galaxy surveys and weak lensing observations, would also provide important constraints on Ho\v{r}ava gravity. This is worthy of further investigations.

\acknowledgments{We thank Debaprasad Maity for his useful
suggestions and comments on the subject. C.-I Chiang wishes to
thank Shu-Heng Shao, Wei-Ting Lin, Kung-Yi Su, Che-Min Shen and
Yeng-Ta Huang for useful and encouraging discussions. Chen is
supported by Taiwan National Science Council under Project No.\
NSC97-2112-M-002-026-MY3, by Taiwan's National Center for
Theoretical Sciences (NCTS), and by US Department of Energy
under Contract No.\ DE-AC03-76SF00515. Gu is supported by the
Taiwan National Science Council under Project No.\ NSC
98-2112-M-002-007-MY3.}


\end{document}